\documentclass[aps,
		prd,
		reprint,
		twocolumn,
		superscriptaddress,
		shortbibliography,
		nofootinbib,
		floatfix,
		notitlepage
		]{revtex4-1}
\pdfoutput=1

\UseRawInputEncoding

\usepackage{slashed}
\usepackage{mathrsfs}
\usepackage{amsmath}
\usepackage{amssymb}
\usepackage{revsymb}
\usepackage{graphicx,accents}
\usepackage{graphicx,epstopdf}
\usepackage{mathrsfs}
\usepackage{bm}
\usepackage{psfrag}
\usepackage{color}
\usepackage{hyperref}
\hypersetup{colorlinks=true,citecolor=Red,urlcolor=Red}
\usepackage{bbm}
\usepackage{bbold}
\usepackage{tabularx}
\usepackage{graphicx}
\usepackage{ulem}
\usepackage[usenames,dvipsnames]{xcolor}

\newcommand{\be}{\begin{equation}}
\newcommand{\ee}{\end{equation}}
\newcommand{\ba}{\begin{array}}
\newcommand{\ea}{\end{array}}
\newcommand{\bea}{\begin{eqnarray}}
\newcommand{\eea}{\end{eqnarray}}
\newcommand{\bd}{\begin{displaymath}}
\newcommand{\ed}{\end{displaymath}}

\newcommand{\trm}[1]{\textrm{#1}}

\newcommand{\ud}{\mathrm{d}}
\newcommand{\LCm}{{\scriptscriptstyle -}} 
\newcommand{\LCp}{{\scriptscriptstyle +}}

\newcommand{\LCperp}{{\scriptscriptstyle \perp}}
\newcommand{\bi}{\begin{itemize}}
\newcommand{\ei}{\end{itemize}}

\makeatletter
\DeclareRobustCommand{\cev}[1]{%
  \mathpalette\do@cev{#1}%
}
\newcommand{\do@cev}[2]{%
  \fix@cev{#1}{+}%
  \reflectbox{$\m@th#1\vec{\reflectbox{$\fix@cev{#1}{-}\m@th#1#2\fix@cev{#1}{+}$}}$}%
  \fix@cev{#1}{-}%
}
\newcommand{\fix@cev}[2]{%
  \ifx#1\displaystyle
    \mkern#23mu
  \else
    \ifx#1\textstyle
      \mkern#23mu
    \else
      \ifx#1\scriptstyle
        \mkern#22mu
      \else
        \mkern#22mu
      \fi
    \fi
  \fi
}

\newcommand*\xbar[1]{%
  \hbox{%
    \vbox{%
      \hrule height 0.5pt 
      \kern0.2ex
      \hbox{%
        \kern-0.15em
        \ensuremath{#1}%
        \kern-0.15em
      }%
    }%
  }%
}

\graphicspath{ {./figures/} }

\begin{document}
\title{Generation of quasi-monoenergetic positron beams in chirped laser fields}
\author{S.~Tang}
\email{tangsuo@ouc.edu.cn}
\affiliation{Department of Physics, College of Information Science and Engineering, Ocean University of China, Qingdao, Shandong, 266100, China}




\begin{abstract}
High energy photons can decay to electron-positron pairs via the nonlinear Breit-Wheeler process when colliding with an intense laser pulse.
The energy spectrum of the produced particles is broadened because of the variation of their effective mass in the course of the laser pulse.
Applying a suitable chirp to the laser pulse can narrow the energy distribution of the generated electrons and positrons. We present a scenario where a high-energy electron beam is collided with a chirped laser pulse to generate a beam of quasi-monoenergetic $\gamma$-photons, which then decay in a second chirped, UV pulse to produce a quasi-monoenergetic source of high-energy electrons and positrons.
\end{abstract}
\maketitle
%
%
\section{Introduction}
%

When a beam of charged particles collides with an intense laser pulse, the spectrum of produced photons, via the process referred to as the nonlinear Compton scattering (NLC)~\cite{ritus1985quantum,dipiazza12}, is sensitive to the shape of the pulse.
Employing a many-cycle laser pulse will lead to outgoing photon spectra similar to those in a monochromatic background~\cite{nikishov64,KingPRA022809}: well-defined harmonic fringes in lightfront momenta and emission angle.
Collision with short laser pulses will lead to a broadening of outgoing particle harmonic peaks~\cite{Boca:2009zz,PRA2011Seipt,mackenroth11} and richer spectral structures: infra-red structure~\cite{Ilderton_2020,king2020Pulse}, asymmetry in emission angle~\cite{Krajewska062102PRA}, and pronounced interference phenomena~\cite{WistisenPRDInterference,ILDERTON2020135410}.
The spectral broadening can be attributed to the inhomogeneous effective mass of the charged particle moving in the intense laser pulse~\cite{kibble64,Kibble1060,PRA022116,PRR013240}: the variation of the effective mass modifies the velocity of the changed particle during the scattering~\cite{PRE29561996,heinzl10b,PRSTAB050703,SeiptPRA033402}.
It is also known, that if one can prescribe the chirp of the laser pulse, that is, a nonlinear dependency on the phase, then this relativistic broadening of  particle spectra can be compensated for, to generate a narrowband source of high energy photons~\cite{PRL074801,SeiptPRA033402,SeiptPRL204802}.

The decay of a probe photon to an electron-positron pair in an intense electromagnetic field, is often referred to as the nonlinear Breit-Wheeler process (NBW)~\cite{breit34,Reiss1962,nikishov64}, and has been measured experimentally in the landmark E144 experiment over two decades ago ~\cite{burke97,bamber99}. The phenomenology of the process has been investigated theoretically in various types of laser field. First in monochromatic ~\cite{nikishov64}, and constant crossed fields~\cite{nikishov64,narozhny69} and more recently in finite laser pulses~\cite{PRD013010,PRD076017,EPJD2020Titov}, as well as two-colour~\cite{PRA052125} and double-pulse fields~\cite{JANSEN201771,Titov:2018bgy,ilderton2019coherent}. As in NLC process, harmonic structure also arises in the outgoing electron-positron pair spectra operating in a many-cycle laser pulse. However, this structure is only clearly discernible when the centre-of-mass energy reaches the threshold of $2mc^2$ already with only a low number of laser photons, where $m$ is electron (positron) rest mass and $c$ is the speed of light.

Analogous to the spectral broadening in the \mbox{NLC} process, the variation of the electron-positron pair's effective mass in the course of the intense laser pulse also induce a broadening in their energy spectra.
The current paper is a proof of principle calculation to show that a suitable nonlinear chirp of a laser pulse can also be employed to counterbalance the spectral broadening in the NBW process, and proposes a simple two-step scenario to provide a quasi-monoenergetic source positrons.
The existence of a quasi-monoenergetic positrons source would be useful in the electron-positron colliding experiments~\cite{PhysRev.124.1577,Myers1990,tang2019one}.


The study of laser chirp's effect on positron spectra, is partly motivated by upcoming high-energy experiments \mbox{LUXE} at \mbox{DESY}~\cite{Hartin:2018sha,abramowicz2019letter} and \mbox{E320} at \mbox{FACET-II}~\cite{Joshi_2018,slacref1}, where photons with energies $\mathcal{O}(10~\trm{GeV})$ are planned to be generated, either directly in the laser pulse through Compton scattering of the electrons (LUXE and E320), or from a separated bremsstrahlung and inverse Compton source (LUXE). The centre-of-mass energy can be effectively increased, and hence the multi-photon harmonic region of the Breit-Wheeler process approached, by generating higher-order harmonics of the interaction laser, using e.g. relativistic plasmas ~\cite{Rodel:2012aa,TangPRE2017,Tang_2019}.

The paper is organised as follows. In Sec.~\ref{Sec_2}, we present the spectrum of produced positrons in the NBW process, and investigate the contributions from the stationary phase points. We then analysis the broadening of the positron spectrum and propose a special laser frequency chirp to counteract the spectral broadening. In Sec.~\ref{Sec_3} we demonstrate numerical implementations of our chirp prescription in narrowing the positron spectra from a single high-energy photon and from the $\gamma$-ray obtained through the NLC process of a high-energy electron. We conclude in Sec.~\ref{conclusion}.

\section{Theoretical framework}~\label{Sec_2}
We consider the scenario in which a high-energy photon with momentum $\ell$ colliding (almost) head-on with a laser pulse produces a pair of electron and positron. The laser pulse is simplified as a plane wave with scaled vector potential $a^{\mu}=eA^{\mu}(\phi)$ and wavevector $k^\mu = \omega_0(1,0,0,1)$ where $\phi=k\cdot x$ and $\omega_0$ is the laser frequency at the initial phase point $\phi_{i}$ at which the laser is turned on.
The interaction energy is characterised by $\eta_{\ell}=k\cdot \ell/m^{2}$.
We use natural units $\hbar=c=1$ throughout and the fine structure constant is $\alpha=e^2\approx1/137$.

The angular-resolved spectrum of the produced positron can be formulated as
\begin{align}
\frac{\ud^{3}\textrm{P}}{\ud s \ud^{2} q_{\LCperp}}&=\alpha\frac{|I|^2+\left(SI^{*}+IS^{*}-2F\cdot F^{*}\right)g}{(2\pi)^2\eta^2_{\ell}(1-s)s}\,,
\label{Eq_NBW_Spectrum}
\end{align}
where $g\equiv[s^2+(1-s)^2]/[4s(1-s)]$. We sum over the spin of the outgoing particles and average over the polarisation of the incoming photon. The spectrum~(\ref{Eq_NBW_Spectrum}) is parameterised by the three components of the positron momentum $p$: these are $s=k\cdot p/k\cdot \ell$, the fraction of the photon light-front momentum taken by the positron, and \mbox{$q_{\LCperp}=(q_{x},q_{y})$}, \mbox{$q_{x,y}= p_{x,y}/m-s\ell_{x,y}/m$}, positron momenta in the plane perpendicular to the laser propagating direction. 
$q_{\LCperp}$ reflects the angular spread of the produced positron around the photon incident direction. In our parameter region, the angular spread is extremely narrow, and for head-on collisions $\ell_{\LCperp}=0$, $q_{\LCperp}\approx \gamma_{p}\theta_{p}(\cos\psi,\sin\psi)$ where $\theta_{p}$ and $\psi$ are the polar and azimuthal angles of the positron, and $\gamma_{p}$ is the positron energy factor.

 The functions $I$, $F$ and $S$ are defined as
\begin{align}
I &= \int \ud\phi \left(1 - \frac{\ell\cdot \pi_{p} }{\ell\cdot p}\right)e^{i\Phi(\phi)}\nonumber\,,\\
F^{\mu} &= \frac{1}{m}\int \ud\phi~a^{\mu}(\phi)~e^{i\Phi(\phi)}\nonumber\,,\\
S  &= \frac{1}{m^2}\int \ud\phi~a(\phi)\cdot a(\phi)~e^{i\Phi(\phi)}\,,\nonumber
\end{align}
with the exponent:
\begin{align}
\Phi(\phi)= \int^{\phi}_{\phi_{i}}\ud \phi'\frac{\ell\cdot \pi_{p}(\phi')}{m^{2}\eta_{\ell} (1-s)}\,,
\label{Eq_exponent}
\end{align}
where $\pi_{p}$ is the instantaneous momentum of the positron in the field:
\begin{align}
\pi_{p}=p+a-\frac{p\cdot a}{ k\cdot p}k-\frac{a^2}{2k\cdot p}k\,.
\end{align}

The completed derivation of the NBW pair production probability could start from an S-matrix element with Volkov wavefunctions~\cite{volkov35} and has been well documented in the literature, see for example~\cite{king2020uniform} for an introduction and~\cite{ILDERTON2020135410} for an analogous presentation for NLC. Because of the charge symmetry, the spectrum~(\ref{Eq_NBW_Spectrum}) can also be applied to the produced electron by just changing the corresponding definitions for the electron. After doing the transverse integral over $q_{\LCperp}$ , the spectrum~(\ref{Eq_NBW_Spectrum}) shows the symmetry: $P(s)=P(1-s)$.

\subsection{Quasi-monoenergetic positron beams from chirped laser pulses}

We now choose, as an example, the vector potential $a^{\mu}$ with circular polarisation:
\begin{align}
a^{\mu}(\phi)&=m\xi \left[0,\cos\Psi(\phi), \sin\Psi(\phi), 0\right] f(\phi)\,,
\end{align}
in which $\xi$ and $f(\phi)$ are the normalised pulse amplitude and envelope, $\ud \Psi(\phi) /\ud\phi = \omega(\phi)/\omega_{0}$ is the chirped frequency of the pulse. At the initial phase $\phi_i=0$, $f(\phi_i)=0$ and $\omega(\phi_i)=\omega_{0}$.

We request (\romannumeral1) that the pulse duration is sufficiently long and the variation of the pulse local amplitude is much slower than the laser frequency, and (\romannumeral2) that the variation of the local frequency $\omega(\phi)$ is on the same time scale as the pulse local amplitude (This will be clear later.): $\omega'(\phi),~f'(\phi)<<1$. Under these conditions, we can then apply the \textit{slowly-varying} approximation that terms of order $f'(\phi)$ [$\omega'(\phi)$] can be neglected~\cite{PRA063110}.

The exponent~(\ref{Eq_exponent}) can be expressed approximately as
\begin{align}
\Phi(\phi)&\approx \kappa(\phi)\phi-\zeta(\phi)\sin\left[\Psi(\phi)-\psi\right]
\end{align}
where
\begin{align}
\kappa=& \frac{\ell\cdot p}{m^{2}\eta_{\ell}(1-s)} +\frac{\xi^2}{2\eta_{\ell}(1-s)s}\frac{1}{\phi}\int^{\phi}_{\phi_{i}}\ud \tilde{\phi} f^{2}(\tilde{\phi})\,,\nonumber\\
\zeta=&\frac{\xi f(\phi)}{\eta_{\ell}(1-s)s}\frac{\omega_{0}}{\omega(\phi)}|q_{\LCperp}|\nonumber\,.
\end{align}
With the Fourier expansions:
\begin{align}
e^{-i\zeta\sin(\Psi-\psi)}&=\sum_{n=-\infty}^{+\infty}J_n(\zeta)e^{in\psi}e^{-in\Psi}\,,
\end{align}
where $J_n(\zeta)$ is the Bessel function of the first kind, the functions $S$ can be expanded approximately as a series of harmonics:
\begin{align}
S & \approx -\xi^{2} \sum_{n=-\infty}^{+\infty}e^{in\psi} \int~\ud\phi~f^2(\phi)~J_n(\zeta)e^{i\Omega(\phi)}\,,
\label{Eq_S_Fourier}
\end{align}
where $\Omega(\phi)=\kappa(\phi)\phi-n\Psi(\phi)$, the harmonic order $n$ means the net number of the laser photons absorbed from the background field. The dependence of the argument $\zeta$ on the pulse envelope $f(\phi)$ indicates that the contribution of each harmonic varies during the course of the pulse and the high order harmonics contribute only at the pulse centre where $f(\phi)\rightarrow1$.  The functions $I$ and $F$ can be calculated in the same way as~(\ref{Eq_S_Fourier}) and obtained with the exactly same exponent term (and the different pre-exponents).

From~(\ref{Eq_S_Fourier}), one can see that the main contribution to the functions $I$, $F$ and $S$, and therefore to the final spectrum~(\ref{Eq_NBW_Spectrum}), comes from the \textit{stationary phase point} where
\begin{align}
\frac{\partial}{\partial \phi}\Omega(\phi)=\frac{q^{2}_{\LCperp}+m^2_{*}/m^2}{2\eta_{\ell}(1-s)s} - n\frac{\omega(\phi)}{\omega_{0}}=0\,.
\label{Eq_NBW_Stationary_cir}
\end{align}
where $m_{*}(\phi)=m[1+\xi^2 f^{2}(\phi)]^{1/2}$ denotes the effective mass of the produced positrons in the laser pulse~\cite{Kibble1060,PRA022116}.

Let us first consider the standard case with constant frequency: $\omega(\phi)=\omega_{0}$. The stationary condition~(\ref{Eq_NBW_Stationary_cir}) implies a chirp in the positron energy varying with the pulse envelop, \mbox{$\ud s/\ud \phi\neq0$}: For the $n$th harmonic, the positron energy is in the spectral region \mbox{$s'_{n,\LCm}(\phi)\leq s \leq s'_{n,\LCp}(\phi)$}, where
\begin{align}
s'_{n,\pm}(\phi) = [1\pm \sqrt{1-2 m^{2}_{*}/(n\eta_{\ell}m^2)}~]/2\,,\label{Eq_chirped_sn}
\end{align}
shifting between the linear [$f(\phi)=0$] and nonlinear [$f(\phi)=1$] BW spectral lines. This shifting stems indeed from the variation of the positron's effective mass $m_{*}(\phi)$ during the course of laser pulse~\cite{SeiptPRA033402,SeiptPRL204802}.

To exclude the energy chirp in the positron spectrum, one simple idea from~(\ref{Eq_chirped_sn}) is to adapt the energy parameter $\eta_{\ell}$ by prescribing the laser frequency with a specific chirp: \mbox{$\eta_{\ell}\sim \omega(\phi)\sim 1+\xi^2f^{2}(\phi)$}, to compensate the variation of the particle effective mass in the laser pulse. As one can see, this frequency chirp is modulated by the field intensity and on the same time scale as the pulse envelope, satisfying the request before.

Based on the stationary condition~(\ref{Eq_NBW_Stationary_cir}), this frequency chirp can be prescribed by solving the differential equation:
\begin{align}
\frac{\ud \omega}{\ud \phi}=\frac{\omega_{0}}{2n\eta_{\ell}(1-s)s}\frac{\ud}{\ud\phi}[q^{2}_{\LCperp}+1+\xi^2 f^{2}(\phi)]\,,
\end{align}
with the initial conditions: \mbox{$f(\phi_i)=0$} and \mbox{$\omega(\phi_i)=\omega_{0}$}, and acquired with its explicit expression:
\begin{align}
\omega(\phi)=\omega_{0}\left[1+ \xi^2 f^{2}(\phi)/(q_{\LCperp}^2+1)\right].
\label{Eq_chirped_frequency}
\end{align}
From~(\ref{Eq_chirped_frequency}), one can see a remarkable fact that this chirp prescription is irrelevant to the harmonic order $n$, which indicates this spectral broadening can be removed from all the harmonic lines at the same time. One should also note that the complete counterbalance of the broadening can only happen at a particular outgoing angle: $|q_{\LCperp}|\sim \gamma_{p}\theta_{p}$ specified by the chirp~(\ref{Eq_chirped_frequency}). For a realistic detector with nonzero angular acceptance $\Delta\theta>0$, the spectral peaks of the probed positrons would shift slightly from the harmonic lines with a finite energy spread (See the results later in Fig.~\ref{Fig_NLC_NBW}).

Similar discussions can also be applied to linearly polarised field backgrounds: \mbox{$a^{\mu}(\phi)=m\xi \left[0,\cos\Psi(\phi), 0, 0\right] f(\phi)$},
and the prescription of frequency chirp~(\ref{Eq_chirped_frequency}) is exactly the same, except that $\xi^{2}\rightarrow\xi^{2}/2$.

\section{Numerical result}~\label{Sec_3}
In this section, we first present a numerical example of a head-on collision between a $13.1\,\trm{GeV}$ photon and a laser pulse with and without the frequency chirp, and then we consider the scenario in which the high-energy photon is replaced with a beam of $\gamma$-photons obtained from the NLC process of a $16.5~\trm{GeV}$ electron.

We are interested in the on-axis positrons collimated in the direction of the incoming photon $q_{\LCperp} \rightarrow0$ for higher yield, and thus apply the frequency chirp:
\begin{align}
\omega(\phi)=\omega_{0}\left[1+\xi^2 f^{2}(\phi)\right].
\label{Eq_chirped_on_axis}
\end{align}
Inserting back into the stationary condition~(\ref{Eq_NBW_Stationary_cir}), one can then get the unchirped positron spectrum peaked at the linear BW harmonic line
\begin{align}
s_{n,\pm} =[1\pm \sqrt{1-2/(n\eta_{\ell})}~]/2
\label{Eq_chirped_linearsn}
\end{align}
at the angle $\theta_{p}=0$. To produce a narrow-band positron beam at the cone angle $\theta_{p}\sim|q_{\LCperp}|/\gamma_{p}$, one can employ the frequency chirp~(\ref{Eq_chirped_frequency}). 

To improve the interaction energy parameter $\eta_{l}$, we employ the laser pulse with the UV frequency $\omega_0=15.5\,\trm{eV}$ and the envelope $f(\phi)=\sin^2(\phi/2N)$ where $0<\phi <2N \pi$ and $N=16$. The (areal) energy of a plane laser pulse can be calculated as:
\begin{align}
E &=-\frac{\omega_0}{4\pi\alpha \lambdabar^{2}_{e}} \int \ud \phi \left(\frac{1}{m}\frac{\ud a}{\ud \phi}\right)^2\,,
\end{align}
in which $\lambdabar_{e}=1/m=386.16~\trm{fm}$ is the electron's reduced Compton wavelength. For circularly polarised laser pulses with the frequency chirp~(\ref{Eq_chirped_on_axis}), one can obtain
\begin{align}
E=\frac{\omega_0 \xi^{2}}{4\alpha \lambdabar^{2}_{e}}  \left(\frac{3}{4}+\xi^2 \frac{35}{32}+\xi^4 \frac{231}{512}\right) N\,.
\end{align}
The first term in the bracket corresponds to the laser pulse with a constant frequency: $\omega(\phi)=\omega_{0}$.

The choice of particle energy parameters is motivated by the upcoming high-energy experiments such as \mbox{LUXE} at \mbox{DESY}~\cite{Hartin:2018sha,abramowicz2019letter} and \mbox{E320} at \mbox{FACET-II}~\cite{Joshi_2018,slacref1}. The strong UV laser pulse can be obtained through the plasma harmonic generation driven by an ultrahigh-power optical laser pulse~\cite{Rodel:2012aa,TangPRE2017,Tang_2019}.
The frequency chirp~(\ref{Eq_chirped_on_axis}) can be plugged into the laser spectrum via the coherent superposition of two linearly and oppositely chirped laser pulses with suitable time delay~\cite{SeiptPRL204802}.

\subsection{NBW}
\begin{figure}[t!!!!]
\center{\includegraphics[width=0.45\textwidth]{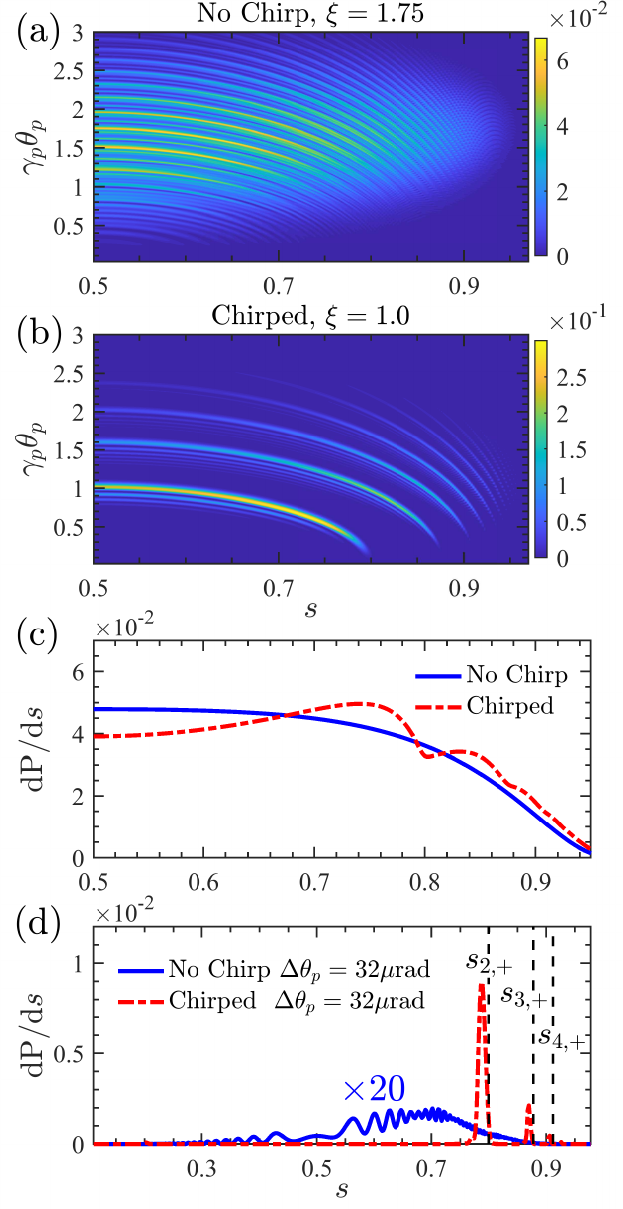}}
\caption{Upper two panels: Angular-energy distribution $\ud^2 \trm{P}/[\ud s~\ud (\gamma_p\theta_{p})]$ of the produced positrons via the \mbox{NBW} process from a high-energy photon ($13.1~\trm{GeV}$) in a circularly polarised laser pulse without (a) and with (b) the frequency chirp. Bottom two panels: Comparison of the positron energy spectra in the laser pulse with (red line) and without (blue line) the frequency chirp within the whole angular spread: \mbox{$\Delta\theta_{p}=\pi$} (c) and a narrow acceptance: \mbox{$\Delta \theta_{p}=32~\mu\trm{rad}$} (d). In (d), the spectral curve for the unchirped case is multiplied by a factor $20$ for visibility. The vertical dashed lines show the location of the first three harmonics in the NBW process: $s_{2,+}=0.80$, $s_{3,+}=0.88$ and $s_{4,+}=0.91$. The chirped laser pulse ($\xi=1$, $\omega(\phi)=\omega_{0}[1+\xi^{2}f^{2}(\phi)]$) has the same energy as the unchirped laser pulse ($\xi=1.75$, $\omega(\phi)=\omega_{0}$).}
\label{Fig_Chirp_NBW}
\end{figure}

Fig.~\ref{Fig_Chirp_NBW} depicts the narrowing of the positron spectra from the chirped laser pulse benchmarked with the results from a constant-frequency pulse. As shown in Fig.~\ref{Fig_Chirp_NBW} (a) and (b), the frequency chirp~(\ref{Eq_chirped_on_axis}) can effectively compensate the broadening of each harmonic line: For the unchirped case in (a), harmonic lines are broadened with plenty of subsidiary peaks and overlap together to be a continuum spectral domain in the positron angular-energy distribution. 
However for the chirped case in (b), the angular-energy distribution is comprised of a number of well-separated harmonic lines.
Around each harmonic line, the distribution is narrowed to be a single peak at the small outgoing angle $\theta_{p}\rightarrow 0$ and is slightly broadened at larger scattering angles with some subpeaks, which comes from the interference between contributions from different stationary points.
At the same time, the amplitude of the harmonic lines in the chirped case are significantly improved.

From Fig.~\ref{Fig_Chirp_NBW} (c), one can see that, with the same laser energy, the chirped laser pulse can produce, approximately, the same number of positrons as the laser pulse with constant frequency. The difference between the positron spectra appears around the harmonic lines $s_{n,\pm}$~(\ref{Eq_chirped_linearsn}), where the harmonic order $n$ must be $\geq2$ because of our parameter setup.
This difference becomes significant in Fig.~\ref{Fig_Chirp_NBW} (d) where the angular-energy distribution is integrated within a narrow angular spread \mbox{$\theta_p<\Delta\theta_{p}/2=16~\mu\trm{rad}$}: the positron spectrum from the chirped pulse background spikes around the harmonic lines $s_{n\geq2,+}$, see the vertical dashed lines in Fig.~\ref{Fig_Chirp_NBW} (d), and has a much higher amplitude than that from the unchirped laser pulse which gives a much lower and broader spectrum in the on-axis direction.
We label the location of the first three harmonics: $s_{2,+}=0.80$, $s_{3,+}=0.88$ and $s_{4,+}=0.91$ in the chirped case.
As one can see, 
the first spectral peak has a much higher amplitude and narrower energy spread $\Delta s/s \approx 1.7\%$ than other higher-order peaks, where $\Delta s$ means the full width at the half maximum of the peak.

As one can see that the $x$-axes of the panels in Fig.~\ref{Fig_Chirp_NBW} (a)-(c) are plotted in the region $0.5<s<1$ because of the symmetry: $\trm{P}(s)=\trm{P}(1-s)$, see~(\ref{Eq_NBW_Spectrum}). However, the spectrum in Fig.~\ref{Fig_Chirp_NBW} (d), which is plotted for the whole spectral region $0<s<1$, shows an asymmetric distribution in the higher energy region $s>0.5$. This is because the lower-energy positrons distribute in a broader angular region: $\gamma_{p}\rightarrow 0$ leading to $\theta_{p}\rightarrow \pi/2$. Therefore, the lower-energy positrons could be simply excluded from the generated high-energy positron beam by an angular selection. 

\subsection{NLC + NBW}
GeV $\gamma$-rays generated by high-energy electron beams via the NLC process has been analysed in detail in~\cite{KingPRA022809,TANG2020135701}. The spectrum of the emitted $\gamma$-rays is presented in~\cite{ILDERTON2020135410} in the same way as~(\ref{Eq_NBW_Spectrum}). With a similar stationary phase analysis, one can prove that a beam of quasi-monoenergetic $\gamma$-photons can be obtained in a well-chirped laser background~\cite{SeiptPRA033402}.
Replacing the seed photon used in Fig.~\ref{Fig_Chirp_NBW} with these acquired high-energy $\gamma$-photons, a source of narrow-band positrons can be obtained. This two-step scenario can be regarded as a part of the trident process~\cite{ilderton11,king13b,king18c} in which only real photons contribute.

\begin{figure}[t!!]
\center{\includegraphics[width=0.45\textwidth]{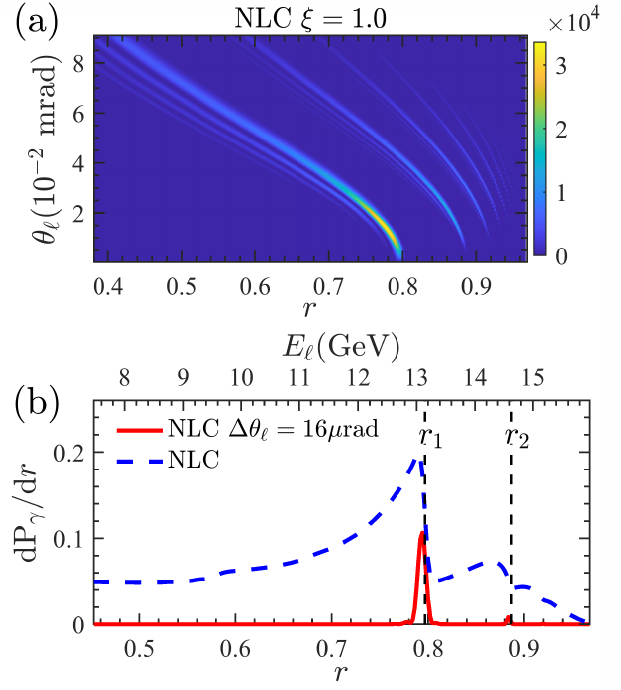}}
\caption{(a) Angular-energy distribution $\ud^2 \trm{P}_{\gamma}/(\ud r~\ud \theta_{\ell})$ of the $\gamma$-photons generated through the NLC process from a high-energy electron $E_{e}=16.5~\trm{GeV}$. (b) Energy distribution of the emitted $\gamma$-photons within the whole angular spread: $\Delta \theta_{\ell}=\pi$ (blue dashed) and a narrow acceptance: $\Delta \theta_{\ell}=16~\mu\trm{rad}$ (red solid). On the top axis is shown the corresponding change in the photon energy $E_{\ell}\approx rE_{e}$ from $7.5~\trm{GeV}$ to $16~\trm{GeV}$. The vertical dashed lines show the location of the first two harmonics in the NLC process: $r_1=0.80$ and $r_{2}=0.89$ corresponding to the photon energy $E_{\ell}=13.1~\trm{GeV}$ and $14.6~\trm{GeV}$. The laser parameters are the same as Fig.~\ref{Fig_Chirp_NBW} (b).}
\label{Fig_NLC_chirp}
\end{figure}

In Fig.~\ref{Fig_NLC_chirp}, we plot the distributions of the emitted photons from a $16.5~\trm{GeV}$ electron head-on 
colliding with the laser pulse parameterized same as in Fig.~\ref{Fig_Chirp_NBW} (b). With the same technique as the NBW case, the frequency chirp can effectively compensate the broadening of the photon spectrum, results in well-separated harmonic lines in the photon angular-energy distribution \mbox{$\ud^{2} \trm{P}_{\gamma}/\ud\theta_{\ell}\ud r$} in Fig.~\ref{Fig_NLC_chirp} (a), especially for small angle scatterings $\theta_{\ell}\rightarrow 0$~\cite{SeiptPRA033402,SeiptPRL204802}, where $r=k\cdot \ell/k\cdot p_{e}$ is the fraction of the light-front momentum taken by the scattered photon from the seed electron, $p_{e}$ is the electron momentum and $\theta_{\ell}$ is the polar angle of the scattered photon.

With a narrow acceptance: $\Delta \theta_{\ell}=16~\mu\trm{rad}$ collimated in the on-axis direction in Fig.~\ref{Fig_NLC_chirp} (b), the collected photons distribute tightly around the first harmonic line \mbox{$r_1=0.80$}, corresponding to the energy \mbox{$E_{\ell}=13.1~\trm{GeV}$}, with the energy spread \mbox{$\Delta r/r\approx 1.3\%$}, and sub-peak around the second harmonic line $r_{2}=0.89$ at $E_{\ell}=14.6~\trm{GeV}$, see the vertical dashed lines in Fig.~\ref{Fig_NLC_chirp} (b), where $r_v=2v\eta_{e}/(2v\eta_{e}+1)$, $\eta_{e}=k\cdot p_{e}/m^{2}$, $v\geq1$ denotes the net number of laser photons absorbed in the NLC process.
The other higher-order subpeaks are too small to view.
With a broader angular selector $\Delta \theta_{\ell}=\pi$, the energy spread of the collected $\gamma$-photons would be much larger, shown as the blue dashed line in Fig.~\ref{Fig_NLC_chirp} (b).

\begin{figure}[t!!]
\center{\includegraphics[width=0.45\textwidth]{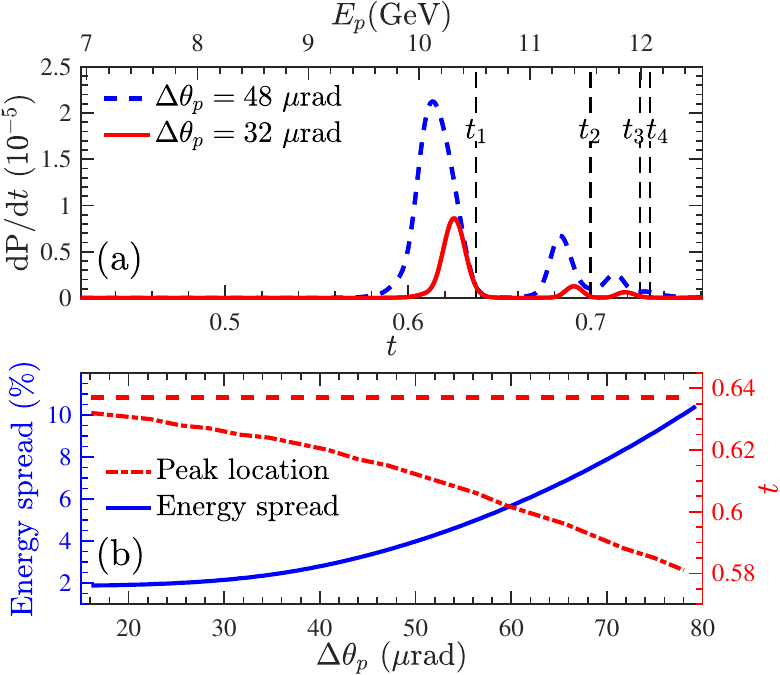}}
\caption{(a) Energy distribution $\ud \trm{P}/\ud t $ of the positrons generated by the on-axis $\gamma$-photons obtained through the NLC process of a high-energy electron $E_{e}=16.5~\trm{GeV}$. The energy region of the produced positron $E_{p}$ is shown on the top axis. The vertical black dashed lines denote the location of each \textit{combined} harmonic: \mbox{$t_{1,2,3,4}=(0.637,~0.700,~0.726,~0.732)$} corresponding, respectively, to the energy $E_{p}= (10.5,~11.5,~12.0,~12.1)~\trm{GeV}$. (b) Energy spread and peak location of the first harmonic peak with the change of the acceptance $\Delta\theta_p$. The horizontal dashed line denotes the theoretical location of the first \textit{combined} harmonic: $t_{1}$. The energy spectrum of the $\gamma$-photons used in the calculation is plotted as the red solid line in Fig.~\ref{Fig_NLC_chirp} (b).}
\label{Fig_NLC_NBW}
\end{figure}

Making use of the obtained $\gamma$-ray spectrum : \mbox{$\rho_{\gamma}(r)=\ud\textrm{P}_{\gamma}/\ud r$} in Fig.~\ref{Fig_NLC_chirp} (b), we can calculate the total number of the generated positrons:
\begin{align}
\textrm{P}&=\frac{\alpha}{(2\pi\eta_{e})^2}\int^{1}_{0}\ud t\int^{1}_{t}\ud r\rho_{\gamma}(r) h(r,t)
\end{align}
where $t=k\cdot p/k\cdot p_{e}$ denotes the fraction of the light-front momentum transferring from the seed electron to the produced positron, and
\begin{align}
h=\int\ud^{2} q_{\LCperp}\frac{|I|^2+(SI^{*}+IS^{*}-2F\cdot F^{*})g}{(r-t)rt}
\end{align}
in which $g\equiv [t^2+(r-t)^2]/4t(r-t)$ is re-defined. 

Here, we ignore the small angular spread of the collected $\gamma$-photons to simplify the numerical calculations. The selection of small-angle photons can be done in experiments by adjusting the separation between the two laser pulses used in each process. The induced numerical errors can be offset by using a much broader acceptance to collect the produced positrons: $\Delta\theta_p\gg\Delta\theta_{\ell}$, see the result in Fig.~\ref{Fig_NLC_NBW}. 

In Fig.~\ref{Fig_NLC_NBW} (a), we plot the yield of the positrons from the on-axis $\gamma$-photons ($\Delta \theta_{\ell}=16~\mu\trm{rad}$) obtained through the \mbox{NLC} process discussed in Fig.~\ref{Fig_NLC_chirp}.
With the acceptance $\Delta\theta_p=32~\mu\trm{rad}$ along the direction of the seed electron, most of the positrons are collected in a narrow energy region peaked at $t=0.625$ with a narrow energy spread $\Delta t/t \approx 2.2\%$ and a much higher amplitude than other subpeaks in the higher energy region.
All of these peaks can be related to the \textit{combined} harmonic lines: $t_{u}=r_{v}s_{n,+}(r_v)$, where $s_{n,+}(r_v)$ is the value of the \mbox{NBW} harmonic line calculated with the \mbox{NLC} harmonic energy $\eta_{\ell}=\eta_{e}r_{v}$, see the vertical dashed lines in Fig.~\ref{Fig_NLC_NBW} (a):
The dominant peak is relevant to the first \textit{combined} harmonic line $t_{1}= r_{1} s_{2,+}= 0.637$ corresponding to the energy $E_{p}\approx 10.5~\trm{GeV}$;
The one appears around $t=0.691$ corresponds to the second \textit{combination} $t_{2}=r_{1} s_{3,+}\approx 0.700$ with the energy $E_p\approx 11.5~\trm{GeV}$, and the one appears around $t=0.719$ may come from the sum of the \textit{combinations}: $t_{3}=r_{1} s_{4,+}=0.726$ and $t_4=r_2s_{2,+}=0.732$.
The red-shift of each peak location is because of the broad acceptance $\Delta \theta_p$ and can be reduced by narrowing the acceptance, see the red dotted-dashed line in Fig.~\ref{Fig_NLC_NBW} (b): the dominant peak moves asymptotically back to the location of the first \textit{combined} harmonic line $t_{1}$ with a decreasing acceptance.

To improve the brilliance of the positron beam, one needs to increase the detector acceptance: with a larger acceptance \mbox{$\Delta\theta_{p}=48~\mu\trm{rad}$} in Fig.~\ref{Fig_NLC_NBW} (a), the amplitude of the spectral peak becomes much higher, and at the same time, its energy spread is broadened to be $3.8\%$.
As shown in Fig.~\ref{Fig_NLC_NBW} (b), the energy spread of the probed positrons increases with the raising of the detector acceptance $\Delta\theta_{p}$.
With a broad acceptance $\Delta\theta_{p}\approx 80~\mu\trm{rad}$, one can acquire a positron beam with the energy spread around $10\%$, and with a relatively narrow acceptance $\Delta\theta_{p}<56~\mu\trm{rad}$, the energy spread of the positron beam can be simply controlled to be less than $5\%$.


\section{Conclusion}~\label{conclusion}

We investigated the nonlinear Breit-Wheeler process in a chirped laser background with intensity $\xi\sim 1$.
Via the standard stationary-phase analysis, we elaborated the broadening of the produced positron spectrum stemming from its inhomogeneous effective mass during the course of the laser pulse, and proved that with a suitable frequency chirp, this broadening can be completely compensated in a specified direction.
We present a proof-of-principle calculation in which a beam of quasi-monoenergetic $\gamma$-photons are obtained from a chirped laser pulse via the nonlinear Compton scattering process and then are used to produce electron-positron pair in the second chirped laser pulse to provide a quasi-monoenergetic source of positrons.
The produced positrons are tightly gathered in a narrow energy region around the \textit{combined} harmonic lines from the relevant processes.
By adjusting the detector acceptance, the energy spread of the obtained positrons can be well controlled.

In our calculations, we ignore the energy spread of the seed electron beam, which would be crucial if it is in the same level of or much broader than the predicted positron energy spread. To obtain the predicted narrow-band positrons, high-quality electron beams with limited energy spread are needed~\cite{slacref1}. We employ the high-power laser pulse with the UV frequency which is critical in our discussion: the UV laser frequency guarantees the scattering of high-energy $\gamma$-photons, and then significantly improve the yield of the electrons and positrons by opening the low-order harmonic channels for the following nonlinear Breit-Wheeler process.

As an outlook, the considered two-step scenario is potential to generate highly polarised positron beams: high-energy $\gamma$-photons scattered by the seed electrons are more probable in the polarisation state parallel to the background field~\cite{TANG2020135701}, and polarised $\gamma$-photons are likely to decay to electron-positron pairs in particular spin states~\cite{SeiptPRA052805}.

\section{Acknowledgments}
The author thanks Dr. A. Ilderton and Dr. B. King for useful discussion and careful reading of the manuscript. The author thanks the support from the Young Talents Project at Ocean University of China.


\providecommand{\noopsort}[1]{}

\end{document}